# Morello: Compiling Fast Neural Networks with Dynamic Programming and Spatial Compression

SAMUEL J. KAUFMAN, RENÉ JUST and RASTISLAV BODIK, Computer Science & Engineering, University of Washington, USA

High-throughput neural network inference requires coordinating many optimization decisions, including parallel tiling, microkernel selection, and data layout. The product of these decisions forms a search space of programs which is typically intractably large. Existing approaches (e.g., auto-schedulers) often address this problem by sampling this space heuristically. In contrast, we introduce a dynamic-programming-based approach to explore more of the search space by iteratively decomposing large program specifications into smaller specifications reachable from a set of rewrites, then composing a final program from each rewrite that minimizes an affine cost model. To reduce memory requirements, we employ a novel memoization table representation, which indexes specifications by coordinates in $Z_{\geq 0}$ and compresses identical, adjacent solutions. This approach can visit a much larger set of programs than prior work. To evaluate the approach, we developed Morello, a compiler which lowers specifications roughly equivalent to a few-node XLA computation graph to x86. Notably, we found that an affine cost model is sufficient to surface high-throughput programs. For example, Morello synthesized a collection of matrix multiplication benchmarks targeting a Zen 1 CPU, including a 1×2048×16384, bfloat16-to-float32 vector-matrix multiply, which was integrated into Google's gemma.cpp.

## 1 Introduction

Implementing high-throughput deep neural networks (DNNs) requires coordinating many decisions across the implementation (e.g., memory layouts [13]) and between high- and low-level intermediate representations (IRs) (e.g., tiling decisions and instruction selection). An ideal program optimizer would jointly, globally optimize all of these decisions, but the product of these decisions forms a space that is too large to exhaustively explore for any non-trivial DNN and IR. For example, Steiner et al. [18] estimate that there are more than $10^{1239}$ implementations of VGG-19 expressible in Halide.

Compilers cope with this combinatorial explosion by splitting the optimization problem into phases, committing to particular decisions (e.g., according to hand-written heuristics) in each phase, such as by first committing to a loop tiling and later to compatible vector instructions [4, 20]. Auto-schedulers instead sample implementations with heuristic search and heuristic reductions of the search space (e.g., limits on loop depth). Both approaches explore only a small fraction of the space, so they require accurate heuristics to find good implementations. As a result, most prior work has focused on improving these heuristics, especially via machine learning [1, 18].

In contrast, our approach explores a larger space of programs than was previously possible, based on two key innovations. First, we developed an IR and associated cost model that form a search problem with an optimal substructure. Our approach recursively decomposes program specifications into smaller specifications reachable from a set of rewrites—for example, tiling rewrites of a matrix multiplication into loops of smaller matrix multiplications—and composes a final program from the rewrites that minimize an affine cost model. Second, we store the memoization table in a spatial data structure, which indexes specifications by coordinates in $Z_{\geq 0}$ and merges identical, adjacent solutions. This reduces memory requirements of our approach by about four orders of magnitude.

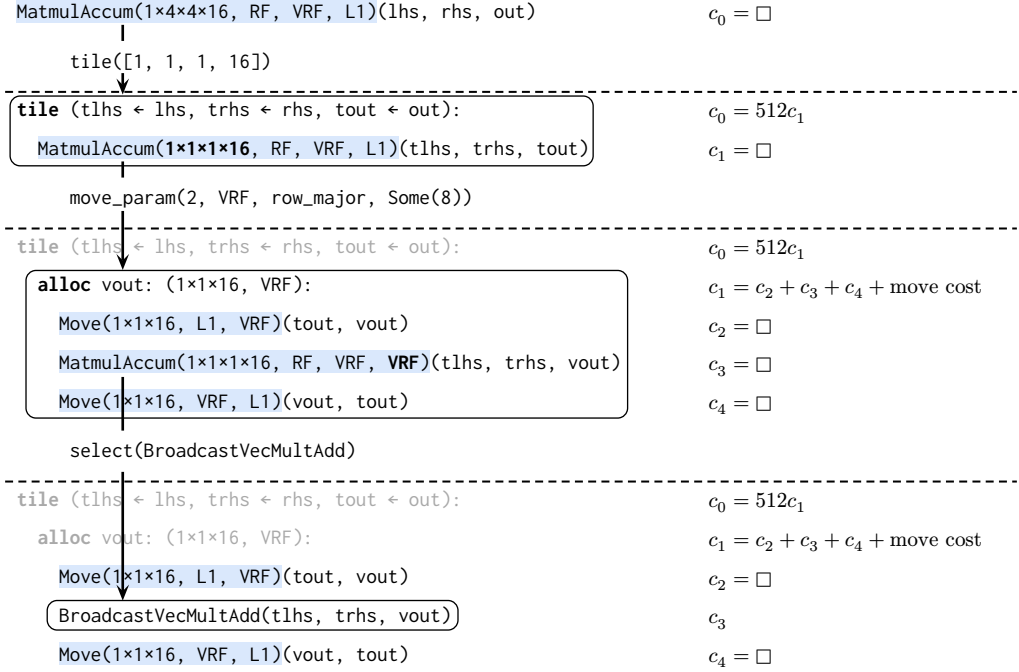

**Figure 1.** An example, partial lowering of a bmkn=$1 \times 4 \times 4 \times 16$ matrix multiplication specification. The specification is tiled to $1 \times 1 \times 1 \times 16$, the output parameter is moved into vector registers, and, finally, the RF-VRF-VRF `MatmulAccum` specification is replaced with the `BroadcastVecMultAdd`. The Moves loading into and storing from the VRF tensor vout are left unimplemented.

To evaluate our approach, we developed Morello, a compiler that lowers specifications roughly equivalent to a few-node XLA computation graph to x86 code via C99 with Clang vector extensions. Notably, we found that an affine cost model is sufficient to surface high-throughput implementations. This is evidenced by success synthesizing a collection of matrix multiplication benchmarks targeting a Zen 1 CPU, including a $1 \times 2048 \times 16384$, bfloat16-to-float32 vector-matrix multiply which was merged into Google's gemma.cpp.

## 2   IR & Scheduling

A high-performance DNN implementation is usually organized as a hierarchy of nested subprograms mapped to the target's memory hierarchy. For example, a GEMM is usually implemented as loops of smaller GEMMs applied to tiles sized to fit particular memory levels [9, 21]. This pattern appears both at the level of individual operators like GEMM and across DNN pipelines. For instance, DeepSpeed-Inference [2] tiles an entire layer normalization block; this involves placing the operators of this block into a loop where each iteration's tiles are in on-chip memory.

Morello's IR models this hierarchical structure. Programs are built recursively by rewriting program specifications into partial programs, which themselves contain specifications of smaller programs. We call this process *scheduling*.

Consider the example in Figure 1, which implements a 4×4×16 accumulating matrix-matrix multiplication.

***Tiling.*** We begin with our goal specification `MatmulAccum(1×4×4×16, RF, VRF, L1)`. Dimensions are ordered batch, $m$, $k$, $n$, and the output is shaped batch × $m$ × $n$. This defines required shapes for the two input and output tensor, as well as the tensors' resident memory levels: registers (RF), vector registers (VRF), and the L1 cache respectively. We first apply the tiling rewrite `tile([1, 1, 1, 16])` to the specification. The result is a loop over the output tile and its dependencies:

```
tile (tlhs: (1×1×1, RF) ← lhs, trhs: (1×1×16, VRF) ← rhs, tout: (1×1×16, L1) ← out):
  MatmulAccum(1×1×1×16, RF, VRF, L1)(tlhs, trhs, tout)
```

The loop consumes zipped tiling iterators over the specification's original tensor arguments, naming the tiles `tlhs`, `trhs`, and `tout`. The body of the loop is a specification with smaller dimensions applied to those tiles, and this smaller specification describes our next optimization problem.

This loop is multi-dimensional, and the iteration order is undefined. Because programs in the Morello IR are built top-down, loops cannot be reordered after construction. Instead, loops can be explicitly ordered along each dimension with a sequence of `tile` rewrites. For example, `tile([1, 4, 1, 16])` followed by `tile([1, 1, 1, 16])`.

***Memory Movement.*** Next, we recurse to rewrite the `MatmulAccum` specification in the loop body. We apply the `move_param(2, VRF, row_major, Some(8))` rewrite, which introduces a move of the output tile (parameter 2) from the L1 cache into the vector register file. (`row_major` and `Some(8)` are discussed in Section 2.4.2 and Section 2.4.1 respectively.)

The result is an `alloc`-binding, a let-binding associated with a move cost (defined in Section 3). `vout` is bound to a new tensor, and the binding's body contains an application of the redex specification with a tensor specification modified to be in VRF. The body also contains two `Move` specifications: a load and a store.

```
tile (tlhs: (1×1×1, RF) ← lhs, trhs: (1×1×16, VRF) ← rhs, tout: (1×1×16, L1) ← out):
  alloc vout: (1×1×16, L1):
    Move(1×1×16, L1, VRF)(tout, vout)  // load from L1 into vector registers
    MatmulAccum(1×1×1×16, RF, VRF, VRF)(tlhs, trhs, vout)
    Move(1×1×16, VRF, L1)(vout, tout)  // store from vector registers into L1
```

In general, a load is inserted when the data movement does not only model hardware cache behavior—specifically, when the destination is software-managed or involves a different layout or data type. Stores are inserted under the same conditions, provided the moved parameter is an output. In this case, load and store `Move`s are added because the faster memory is software-managed and the parameter is an output parameter. Our modeling of memory, including our treatment of hardware caches and tensor layouts (e.g., `row_major` in the above rewrite), is discussed further in Section 2.4.

Notice that a consequence of building programs top-down is that these multiple, nested specifications are optimized independently. Their costs and consequently expected performance do not interact.

***Microkernel Selection.*** Finally, we complete the example from Figure 1 by rewriting not only the nested `MatmulAccum` specification but also the `Move` specifications. We complete the implementation with `select(BroadcastVecMultAdd)` and `select(VectorMove)` rewrites:

```
tile (tlhs: (1×1×1, RF) ← lhs, trhs: (1×1×16, VRF) ← rhs, tout: (1×1×16, L1) ← out):
  alloc vout: (1×1×16, VRF):
    VectorMove(tout, vout)
    BroadcastVecMultAdd(tlhs, trhs, vout)
    VectorMove(vout, tout)
```

At this point, the implementation is complete: no specifications remain and all leaves are micro-kernel applications. During code generation, a microkernel application lowers to one or more individual instructions, such as calls to elementwise multiplication between vectors, or calls to (potentially inlined) functions. In this case, `BroadcastVecMultAdd` lowers to **vbroadcastss** and **vfmadd213ps** instructions, and `VectorMove` lowers to **vmovups** instructions.

The rewrites we have seen so far—tiling, memory movement, and microkernel selection—are sufficient to express a variety of important optimizations: strip-mining is simply tiling, vectorization is the selection of a vectorized microkernel, layout packing is a `move` from one layout to another, and loop fusion is tiling applied to a composed specification rather than one of its components.[1]

## 2.1 Pipelines

Specifications can be composed to express a neural network pipeline. For example, the following is an application of a composition of a producer 1×64×1024×32 matrix multiplication with a consumer 1×64×2048×128 matrix multiplication:

```
Compose(
  [Matmul, Matmul],
  [(1×32×128, GL), (1×64×1024, GL), (1×1024×32, GL), 1×64×128, GL)]
)(a, b, c, out)
```

A composed specification binds its component primitives from outermost to innermost, with each primitive's output passed as the first input to the next, while the remaining inputs are read left-to-right. The rightmost parameter is the final output parameter.

Compositions can be tiled and have their parameters moved, just as primitive specifications can. For example, we can tile and move the output tensor of the above specification:

```
tile (ta: (1×32×32, GL) <- a, tb: (1×32×1024, GL) <- b, tc: (1×32×32, GL) <- out):
  alloc resident: (1×32×32, L1) <- tc:
    Compose(
      [Matmul, Matmul],
      [(1×32×128, GL), (1×32×1024, GL), (1×1024×32, GL), (1×32×32, L1)]
    )(ta, tb, c, resident)
```

Tiling a composition, as opposed to tiling each of its individual primitives, produces a fused loop. It is equivalent to fusing the tiled loops of the individual primitives.

Ultimately, compositions are lowered to their separated primitives via `bufferize` rewrites. These rewrites "split" a composition into producer and consumer parts, which write into and read from a new, intermediate tensor. For example, applying `bufferize(0, L1, row_major, None)` to the nested composition above produces:

---

[1]Morello does not control loop unrolling. While it is an important optimization, we've found that LLVM's unrolling heuristics are sufficient. Morello only unrolls loops when needed to refer to the individual names of a distributed tensor in target code.

```
tile (ta: (1×32×32, GL) <- a, tb: (1×32×1024, GL) <- b, tout: (1×32×32, GL) <- out):
  alloc resident: (1×32×32, L1) <- tout:
    pipeline (intermed: (1×32×32, L1)):
      Matmul((1×32×1024, GL), (1×1024×32, GL), (1×32×32, L1))(tb, c, intermed)
      Matmul((1×32×32, L1), (1×32×32, GL), (1×32×32, L1))(intermed, ta, resident)
```

This new program contains a *pipeline*, which is a node for sequential execution. Pipelines differ from plain blocks in that they imply interlocking lifetimes for intermediate tensors: intermediates live only during the execution of their producing or consuming sub-program.

## 2.2 Spec Language

To this point, we have used a simplified specification language. In fact, the Morello specification language is more expressive. We call an expression in this language a Spec, which has five fields: (1) a type, (2) shape, (3) list of tensor specifications, (4) serial-only flag, and (5) per-level peak memory bounds. The serial-only flag requires an implementation to use at most one hardware thread. A *tensor specification* has five fields: (1) data type, (2) memory level, (3) data layout, (4) memory alignment, and (5) contiguousness.

With these additional fields, we could extend the goal specification of Figure 1 as follows:

```
(MatmulAccum(
  1×4×128×16,
  (bf16, L1, row_major),
  (bf16, L1, col_major, ua, c0),
  (f32, VRF, row_major, v8),
  serial),
[64, 1024, 32768, 1073741824])
```

In this syntax:
- `ua` means "unaligned"; if `ua` is omitted, the tensor is assumed to be aligned;
- `c0` means that tensor values are not guaranteed to be adjacent in memory; if `c0` is omitted, values are assumed to be adjacent;
- `v8` means that each 8-value strip of values shares a vector register; and
- `[64, 1024, 32768, 1073741824]` bounds the peak memory (in bytes) that a satisfying implementation can allocate in each level of the memory hierarchy. These levels are ordered: scalar register file, vector register file, L1 cache, and "global" memory.

Section 2.4 expands on these features.

A Spec is either one of 21 primitives, such as matrix multiplication, convolution, softmax, and memory movement, or it is a composition of these primitives. Twelve of those primitives are paired in that they have both an accumulating and non-accumulating variant such as `MatmulAccum` and `Matmul`. It is usually possible to rewrite a non-accumulating Spec to an output-initializing sub-Spec followed by its accumulating variant.

`Move` Specs unify data movement, memory layout, and upcasting operations, which are typically distinct in other systems. Specifically, a `Move` describes one or more of:

1. Loads and stores when the memory levels of parameters differ. For example, `(Move(4×4, (u8, L1, row_major), (u8, RF, row_major), serial), ...)` is a load from L1 into registers.
2. Data packing when parameter layouts differ. A `(Move(4×4, (u8, L1, row_major), (u8, L1, col_major), serial), ...)` copies data from a tensor with a row-major layout into a tensor with a column-major layout.

3. Type casts when data types differ. A `(Move(4×4, bf16, RF, row_major), (f32, RF, row_major), serial), ...)` copies the bfloat16 value into a tensor with float32 values, upcasting.

## 2.3 User Scheduling

While the Morello IR is designed to support automatic optimization, rewrites are also exposed to programmers via a Halide-style scheduling language [16] embedded in Rust. For example, the above program is generated by the following schedule:

```
let goal = spec!(MatmulAccum(1×4×4×16, RF, VRF, L1), ..);
let implementation = goal
    .tile(&[1, 1, 1, 16])
    .move_param(2, VRF, row_major, Some(nz!(8u32)))
    .subschedule(&[0], |load_spec| load_spec.select(VectorMove))
    .subschedule(&[1], |matmul_spec| matmul_spec.select(BroadcastVecMultAdd))
    .subschedule(&[2], |store_spec| store_spec.select(VectorMove));
```

Rewrites are applied to the leaves of partial implementations. When the program has only one nested SPEC, the target of a rewrite is unambiguous. This is the case for the above `tile` and `move_param` rewrites. When a program node has multiple children, the rewrite target is ambiguous. This is the case for the implementation resulting from `move_param`, which contains three sub-specifications: a `Move` loading data into `VRF`, a nested `MatmulAccum`, and a `Move` storing data from `VRF` back into `L1`. In this case, the programmer chooses the rewrite's target with the `subschedule` operator. In the above example, child SPECs indexed 0 and 2 are replaced with the `VectorMove` microkernel and the child SPEC indexed 1 is replaced with `BroadcastVecMultAdd`.

Rather than manually completing an implementation, programmers may apply some rewrites before completing the partial implementation with synthesis. For instance, a programmer may choose to apply `tile`, `move_param`, and a single `select` rewrite but have Morello synthesize the `Move` SPECs:

```
let db = todo!("Allocate a memoization table");
let goal = spec!(MatmulAccum(1×4×4×16, RF, VRF, L1), ..);
let implementation = goal
    .tile(&[1, 1, 1, 16])
    .move_param(2, VRF, row_major, Some(nz!(8u32)))
    .subschedule(&[1], |matmul_spec| matmul_spec.select(BroadcastVecMultAdd))
    .synthesize_all(&db, None);  // Synthesize the two Moves
```

We expect programmers will often schedule the main arithmetic loops themselves while leaving the implementation of Moves to the synthesizer. To support this workflow, we introduce default child selection: when a rewrite rule has multiple children and would normally require `subschedule` to specify the target SPEC, the rule instead applies to the sole non-Move child—if exactly one exists:

```
let goal = spec!(MatmulAccum(1×4×4×16, RF, VRF, L1), ..);
let implementation = goal
    .tile(&[1, 1, 1, 16])
    .move_param(2, VRF, row_major, Some(nz!(8u32)))
    .select(BroadcastVecMultAdd)  // Schedules child 1 (the arithmetic Spec)
    .synthesize_all(&db, None);
```

## 2.4 Memory Movement

Taking full advantage of modern processors requires carefully scheduling memory movement across the memory hierarchy. Optimal performance requires scheduling register spills and cache misses to minimize memory access and overlap with arithmetic, as well as accurately estimating the costs of layout changes (e.g., packing) and weighing them against the benefits of faster microkernels. This is especially important for unusual tensor shapes.

Morello supports these optimizations by explicitly modeling both hardware- and software-controlled memory movement. Available memory is tracked by tightening the memory bounds of child SPECS during rewrites, preventing spills and unexpected cache misses. In addition, child SPECS' tensor specifications are updated to model tensor layout, contiguousness, and alignment, which affect both the applicability of microkernels and other rewrites as well as the cost of memory movement (e.g., layout affects the number of cache lines).

The alloc-binding and pipeline nodes define clear lifetimes for introduced tensors, so Morello tracks per-level memory consumption by lowering the corresponding memory bounds on child SPECS: subtracting bytes consumed and snapping to the lower power of two (or zero). Morello models memory consumption as the sum of the bytes backing every logical tensor alive, ignoring loop iterators and fragmentation. Morello expects the LLVM backend to hide any loop iterator spills via unrolling and instruction scheduling, and we have found memory fragmentation to be tolerable because DNNs tend to make large, power-of-two-sized allocations.

### 2.4.1 Tensors

A Morello tensor is a multi-dimensional array view into memory: spanning one or more memory buffers or registers. Tensors can be tiled into new tensors over the same data. A tensor has a tensor specification, and tensor specifications abstract over both tiles and tensors backed directly by a buffer. As a result, any program that can be applied to a non-tile tensor can also be applied to a tile with the same tensor specification. This abstraction allows, for example, the optimal implementation of a Matmul SPEC to be reused in the body of a larger Matmul tiled to the same size.

Tiles are additionally equipped with *logical indexing expressions* for each dimension. These expressions map concrete tile and point indices from the tile's coordinate space to the source tensor's coordinate space. (For non-tiling tensors, these expressions are the identity map.)

Tensors backed by memory in register files are distributed across registers and, in the context of a parallel loop, across private memories. Morello provides some control of the tensor-to-register mapping when tensors are assigned to a vector register file via the vector_size field, which selects among aliased register names by specifying the number of values per vector register (e.g., v8).

### 2.4.2 Layouts

Every tensor specification is associated with a *layout*. A layout is a function, which maps a tensor shape to a *buffer indexing expression*, which is a bijection from tensor coordinates to buffer offsets.

Morello supports layouts, which map logical tensor dimensions to regularly nested physical dimensions (e.g., column-major: $mj + i$) with, optionally, a strip dimension of size $s$ (e.g., $m \left\lfloor \frac{j}{s} \right\rfloor + si + (j \bmod s)$). This includes common and important layouts like row-major, NCHW, and packed layouts such as Goto-style matrix packing [9] or NCHWc [8]. Additionally, strips may be odd-even (Faro) shuffled. For example: $m \left\lfloor \frac{j}{s} \right\rfloor + si + \sigma(s, j \bmod s)$, where

$$\sigma(s, x) = 2 \left( x \bmod \left\lfloor \frac{s}{2} \right\rfloor \right) + \left\lfloor \frac{2x}{s} \right\rfloor$$

This allows for lane shuffling between vector registers when the vector size is $\frac{s}{2}$.

During code generation, array indexing expressions are generated by recursively composing the stack of tiles' logical indexing expressions, terminating with the buffer indexing expression of the final tensor's layout.

### 2.4.3  Contiguousness & Alignment

As mentioned in Section 2.2, tensor specifications have *contiguousness* and *alignment* fields, which model whether a tensor's values are adjacent in memory and whether the tensor's memory offset is a multiple of a target-specific offset (in the case of our x86 target, the size of a cache line). These fields affect the costs of memory movements and applicability of certain microkernels.

Both fields store abstract values, which are initialized to a constant $\top$ value for new tensors backed by fresh buffers, at which point the tensor is always both contiguous and aligned. The values are updated according to transfer functions during tiling. Morello supports any analyses, which fit this framework so long as the values can be mapped to flags indicating whether or not the tensor is contiguous/aligned.

Morello's x86 target abstracts contiguousness as the number of inner physical dimensions for which values are adjacent to each other in the underlying buffer. A tensor is said to be contiguous if all physical dimensions are counted. Alignment is abstracted as a boolean, and the transfer function only determines that a tensor is aligned if it was previously aligned and contiguous and if every suffix of physical dimensions is a multiple of the cache line size.

### 2.4.4  Memory Bounds

Memory bounds are updated whenever a buffer is introduced, which happens in two cases: when introducing a `move`, which allocates a buffer for an operand at a new memory level, and when a `pipeline` is introduced, which allocates intermediate buffers between children.

Usually, memory bounds are represented as a data structure mapping memory levels to available bytes. However, when a `Pipeline` is introduced, memory bounds carry additional information about how much additional memory is consumed by the intermediate buffers. If a SPEC in a `Pipeline` is replaced with another `pipeline`, the child's bounds can be updated to take advantage of the fact that their output buffer need not be live for its entire duration.

## 3  Cost Model

Our goal is to maximize generated code's throughput. To that end, Morello's cost model is designed as a proxy for cycles per iteration of the implementation $p$ in a loop with no carried dependencies.

The model is defined compositionally over the program tree. The cost of each program node is an affine function that is no less than the costs of its sub-nodes. This gives our synthesis problem an optimal substructure: selecting an optimal rewrite amounts to selecting the rewrite that yields the lowest-cost program. Concretely, the cost $c(p)$ of a program $p$ is:

$$c(p) = \begin{cases} \left\lceil \frac{\text{trips}}{\text{threads}} \right\rceil \cdot c(\text{body}_p) + s & \tau_p \in \{\text{Loop}\} \\ \sum_{q \in \text{children}(p)} c(q) & \tau_p \in \{\text{Block, Pipeline}\} \\ m(\text{src}) + m(\text{dest}) + \sum_{q \in \text{children}(p)} c(q) & \tau_p \in \{\texttt{alloc-binding}\} \\ k_p & p \text{ is a microkernel} \end{cases}$$

$$m(t) = \text{lines}(t) \cdot \text{weight}_{\text{level}(t)} \cdot ([t \text{ is contiguous}] + 1)$$

Serial loops cost as much as their bodies multiplied by trip count, and parallel loops are the same, adjusted by a target-specific parallelism factor ("threads") and with an additional constant synchronization cost $s$. Moves add costs $m(t)$ proportional to the number of cache lines accessible by and the speed of the memory level of $t$'s backing memory, adjusted according to $t$'s contiguousness. Blocks cost the sum of their child node costs, and microkernels have costs $k_p$.

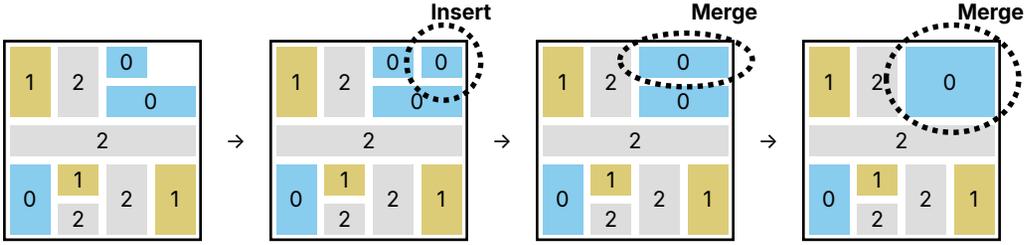

**Figure 2.** Morello's database tables contain space-filling rectangles describing the optimal decision or cost for the SPECS covered by that space. Values are inserted into Morello's R*-Tree-backed database as a unit rectangle that is then merged with adjacent rectangles. The result is a database with many fewer entries.

We chose kernel costs $k_p$ for each $p$ by measuring reciprocal throughput on llvm-mca. We fit $s$ and weight$_l$ through a limited hyperparameter evaluation. Future work could, if necessary, fit parameters in a more principled way, but we have found this cost model is sufficient to rank high-throughput implementations on x86.

## 4  Synthesis

Morello implements a top-down search that computes the optimal implementation for a given SPEC. If the memoization database contains a rewrite for that SPEC, the search procedure returns that rewrite. Otherwise, the procedure evaluates rewrites applicable to the goal SPEC, recurses on that rewrite's SPEC dependencies, and memoizes and returns the rewrite yielding the lowest-cost implementation.

Search explores all applicable rewrites except non-power-of-two tile sizes. Additionally, only the following layouts are considered:

- all regularly nested logical dimension orders, and
- each of those orders extended with at most one packed dimension per logical dimension in any order, where the packed dimension may or may not be odd-even interleaved.

Only single-dimension tile sizes are visited; multi-dimensional tilings are synthesized through recursive tilings.

### 4.1  Memoization Database

Synthesis can quickly compute a large number of optimal scheduling decisions, but storing those decisions becomes difficult as problem sizes approach those of real DNNs. For each SPEC, Morello stores the optimal rewrite decision and its cost, depth, and memory consumption, which requires 11 bytes per entry. Storing a database of 2048-square matrix multiplications and its dependencies requires several petabytes if done naively (approximately $2 \times 10^{14}$ SPECS at 11 bytes each).

Fortunately, optimal decisions and costs are highly regular: the optimal implementation of one SPEC is often identical to that of a SPEC with a slightly varied shape or memory limits. For example, the optimal size-128 and size-64 matrix multiplications may both be implemented as loops around the same microkernel. In this case, the stored rewrite for each of these matrix multiplications is `tile(k)` where `k` is the microkernel shape. Additionally, after normalizing costs to the volume of the SPEC, identical scheduling decisions often have the same normalized cost.

The design of our database exploits these regularities of optimization decision and cost by mapping decisions to coordinates in $Z_{\geq 0}$ such that adjacent entries are likely to match. Memory limit and tile sizes are restricted to powers of two, so adjacent entries along those dimensions are double the size or limit.

Decisions and costs (separated into main cost, depth, and memory consumption components) are inserted into separate R*-Trees as $n$-dimensional unit rectangles. Adjacent rectangles are then greedily merged wherever they represent the same value and together fill their combined, merged bounding box. This is illustrated in Figure 2.

We find that, on average, rectangles store approximately 4,000 values, corresponding to a same-factor reduction in storage requirements.

## 5  Related Work

### 5.1  Fireiron

The Morello IR is most directly inspired by Hagedorn et al.'s Fireiron [12], which does not address synthesis but does explicitly represent memory movement and decomposes program specifications into smaller sub-specifications at each scheduling step. Fireiron demonstrates that syntactic, top-down expansion of specifications into partial programs with nested sub-specifications is a simple and expressive way to schedule tensor programs for GPUs.

### 5.2  DP-Style Optimization

Scheduling high-performance code with dynamic programming is not a new idea, especially in the scientific computing and databases communities. For instance, FFTW [6, 7] is popular collection of high-performance Fast Fourier Transform implementations synthesized by composing pre-written, fixed-size codelets (i.e., microkernels) to minimize a cost with optimal substructure. Likewise, relational database management systems map a logical query over relations to a program over concrete data structures called a plan. The seminal System R query optimizer [17] synthesized plans with bottom-up dynamic programming, and many contemporary, state-of-the-art database systems including Volcano [10] and its more modern incarnation Cascades [11] still use this basic approach.

Among work on DNN optimization, dynamic programming appears in work on an orthogonal problem: scheduling pipeline parallelism. PipeDream [15], which schedules mini-batch pipeline parallelism across accelerators, is an early and influential example. Later systems like DAPPLE [5] and AdaPipe [19] extended this basic DP approach to optimize joint weights/activations- and pipeline-parallel execution and adaptive recomputation, respectively.

### 5.3  Auto-Scheduling Neural Networks

There is a large body of work on auto-scheduling neural networks, which treats optimization as a sequential decision-making problem over an action space of the Halide/TVM scheduling language or other semantics-preserving rewrites. Most use a learned cost model to guide a heuristic search. Examples include the Halide auto-schedulers by Adams et al. [1] targeting small DNNs (ResNet-50 and Conv+ReLU blocks) on x86 and Anderson et al.'s [3] targeting GPUs, as well as TVM's Ansor [22], FlexTensor [23], and Steiner et al.'s [18] deep value iteration-based approach.

No other DNN auto-scheduler collapses the search space by assuming an optimal substructure. However, our key idea of indexing programs by specification is compatible with prior work. For instance, a value iteration approach like that of Steiner et al. could approximate the value of a specification rather than a schedule prefix.

## 6  Future Work

***Nonlinear Computation Graphs.*** Morello supports pipelines but not arbitrary computation graphs. Future work will extend SPEC expressions to support DAGs, specifically extending SPEC composition and generalizing the `bufferize` rewrite to insert intermediate tensors at graph edges. This will enable Morello to express common non-pipeline structures, such as residual connections.

***Pipeline Parallelism.*** Morello's costs cannot accurately model pipeline parallelism, which may lead to inaccuracies on architectures where latency hiding (e.g., offloading) is important. We intend to experiment with compositional synthesis of Pareto frontiers of costs (trading off separate arithmetic and memory latency costs).

***Top-Down Partial Scheduling.*** Finally, it is currently not possible to constrain sub-trees or leaves without first selecting their ancestors. For example, a user cannot synthesize an implementation that uses a specific microkernel without manually scheduling the entire implementation. We intend to allow the user to filter microkernels when synthesizing.

## 7  Conclusion & Results

We introduced a novel approach to synthesizing neural networks built on two key innovations: (1) a compositional IR and cost model that supports dynamic-programming-based synthesis, and (2) a novel application of a spatial database to improve the scalability of our synthesizer beyond what would otherwise be possible.

We implemented this approach and developed a compiler named Morello, which successfully synthesized a collection of matrix multiplication benchmarks targeting a Zen 1 CPU. One of these, a 1×2048×16384, bfloat16-to-float32 vector-matrix multiply, improved on the performance of, and has been integrated into, Google's gemma.cpp [14]. Another, a single-threaded float32, 2048-square matrix multiplication, reaches the theoretical peak performance of the core.

These successes demonstrate scalability and validate our design choices in two ways. First, our set of rewrites is sufficient to express important optimizations for at least matrix multiplication kernels. Second, the affine cost model is sufficient to correctly rank high-throughput implementations. We initially hypothesized that hardware pipelining would pose a problem in the form of overly pessimistic, additive cost composition. However, we have not observed high-throughput implementations being incorrectly ranked for that reason. Building programs strictly top down in our IR is not overly restrictive.

The Morello compiler is under development and publicly available at: https://github.com/samkaufman/morello.

## Acknowledgements

This work has been supported in part the National Science Foundation awards ITE−2132318, CCF−2122950, ITE−2029457, ITE−1936731, CCF−1918027, IIS−1924435, the Intel and National Science Foundation joint research center for Computer Assisted Programming for Heterogeneous Architectures (CAPA NSF CCF−1723352), the CONIX Research Center, one of six centers in JUMP, a Semiconductor Research Corporation (SRC) program sponsored by DARPA CMU 1042741-394324 AM01, a grant DARPA FA8750−16−2−0032, as well as gifts from Adobe, Facebook, Google, Intel, and Qualcomm.